\documentclass[%
 reprint,
 amsmath,amssymb,
 aps,
]{revtex4-1}

\usepackage{graphicx}
\usepackage{dcolumn}
\usepackage{bm}
\usepackage{graphicx}
\usepackage{dcolumn}
\usepackage{bm}
\usepackage{bigints}
\usepackage{color}
\usepackage{amsmath,amssymb,graphicx}
\usepackage{float}
\usepackage{braket}
\usepackage{amsmath}
\usepackage{textcomp}

\begin{document}

\preprint{APS/123-QED}

\title{Statistics of mode area for transverse Anderson localization in disordered optical fibers}

\author{Behnam Abaie}
\author{Arash Mafi}%
 \email{mafi@unm.edu}
\affiliation{Department of Physics \& Astronomy, University of New Mexico, Albuquerque, NM 87131, USA
             \\Center for High Technology Materials, University of New Mexico, Albuquerque, NM 87106, USA}%

\date{\today}

\begin{abstract}
We introduce the mode area probability density function (MA-PDF) as a powerful tool to study transverse 
Anderson localization (TAL), especially for highly disordered optical fibers.
The MA-PDF encompasses all the relevant statistical information on TAL; it relies solely on the physics 
of the disordered system and is independent of the shape of the external excitation. 
We explore the scaling of MA-PDF with the transverse dimensions of the system and show that it 
converges to a terminal form for structures considerably smaller than those used in experiments, hence 
substantially reducing the computational cost to study TAL.
\end{abstract}

\pacs{Valid PACS appear here}

\maketitle
Anderson localization (AL)~\cite{Anderson1,EAbrahams,SegevNaturePhotonicsReview} has been explored in great detail over the years 
in various classical wave ~\cite{Weaver,Graham,dalichaouch1991microwave,John1,John2,John3,Anderson2,Lagendijk1,Hu,Chabanov} and 
quantum systems~\cite{Billy,Lahini2,Lahini3,Abouraddy,Thompson}. In his 1977 Nobel lecture, P.~W.~Anderson declared that AL
``has yet to receive adequate {\em mathematical} treatment'', and ``one has to resort to the indignity of {\em numerical} simulations to 
settle even the simplest questions about it.'' There has since been a great progress in the {\em theoretical} understanding of AL; 
however, theory provides general guidelines while details are left to the {\em numerics}. The necessity of numerical studies  
is more pronounced for AL of electromagnetic waves because a full theoretical treatment is yet to be developed.
For example, the vectorial nature of light~\cite{Skipetrov} and the gradient term of the spatially varying 
permittivity~\cite{schirmacher2017right} can result in drastically different localization properties than one obtains from a 
simpler Schr\"{o}dinger-like equation, which has received the most extensive theoretical treatments.

Although microprocessor clock speed has increased by four orders of magnitude since 1977, numerical studies of AL can still be 
formidable even on supercomputers. In this Letter, we apply detailed numerics together with arguments on scaling
to study AL in a quasi-two dimensional (2D) setting. While even at the quasi-2D level the computational problem is difficult to tackle, 
our novel approach based on scaling makes it doable on a modest computer cluster and provides a viable pathway to expanded studies of AL in 
three dimensions. The problem under study is of practical importance as will be described later.
We study TAL~\cite{Abdullaev,DeRaedt} as introduced by De Raedt, et al.~\cite{DeRaedt}. 
They considered a dielectric waveguide with a transversely random and longitudinally uniform refractive index profile (quasi-2D). 
An optical field that is coupled to this waveguide freely propagates in the longitudinal direction, but
remains fully confined in the transverse plane due to TAL after an initial expansion~\cite{DeRaedt,Schwartz,Lahini,Martin,SalmanOL,SalmanOPEX,SalmanOMEX,MafiAOP}.

The traditional study of TAL is based on launching a beam and analyzing its propagation along the waveguide 
using the beam propagation method (BPM)~\cite{DeRaedt,SalmanOPEX}. It is hard to ensure that the conclusions are not biased
by the choice of the initial launch condition. Moreover, the inner mechanisms of the localization are hidden to the observer.
Here, we propose a study based on the mode-area (MA) probability density function (PDF). The MA-PDF 
encompasses all the relevant statistical information on TAL; it relies solely on the physics of the disordered system and the physical 
nature of the propagating wave, and is independent of the beam properties of the external excitation. It also makes the TAL 
behavior very transparent to the observer, where one can clearly differentiate between the localized and extended modes~\cite{BoundaryPaper,PhysRevB.94.064201}. 
As such, one can strategize the optimization of the waveguide using desired objective functions. These attributes make the modal-based computation 
of the MA-PDF superior to other methods to analyze TAL.

A key observation presented in this Letter is that the MA-PDF can be reliably computed from structures with substantially smaller
transverse dimensions than the practical waveguides used in experiments. In fact, it is shown that the shape of the MA-PDF rapidly 
converges to a terminal form as a function of the transverse dimensions of the waveguide. This peculiar scaling behavior
observed in MA-PDF is of immense practical importance in the design and optimization of such TAL-based waveguides, because one
can obtain all the useful localization information from disordered waveguides with smaller dimensions, hence substantially 
reducing the computational cost. The observed behavior has broad implications in the study of disordered systems, because
it clearly illustrates that the statistical behavior of the large and often computationally formidable disordered problems
in the localization regime can be tackled by looking into smaller systems. 

\begin{figure}[t]
\centerline{
\includegraphics[width=\columnwidth]{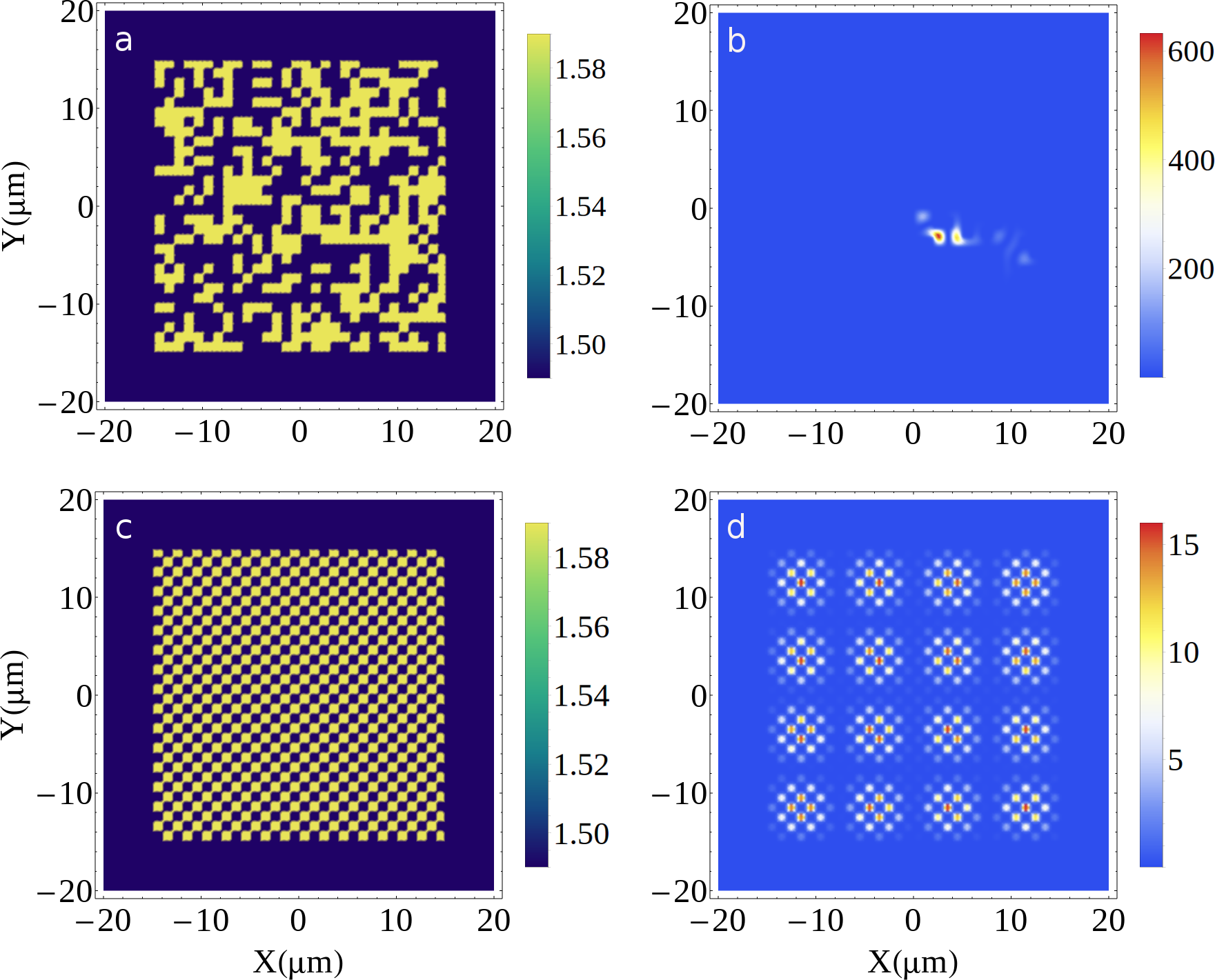}
}
\caption{\label{fig:ModeProfile}
Sample refractive index profiles of (a) disordered  and (c) ordered optical fiber with an arbitrarily selected square cross section. (b), and (d) represent the 
typical calculated modes for the index profiles in (a), and (c), respectively. For a disordered optical fiber modes are transversely localized as opposed to the 
ordered fiber where the mode covers the entire cross section of the fiber.}
\end{figure}

The conclusive report of TAL by De Raedt, et al. is based on the structure sketched in Fig.~\ref{fig:ModeProfile}, where the 
transverse dimension of the waveguide is divided into square pixels of width $d$, where each pixel is randomly chosen to have a 
refractive index value of $n_1=1.0$ or $n_2=1.5$ with equal probabilities. TAL has since been experimentally confirmed by various groups in different
implementations over the past decade~\cite{Schwartz,Lahini,Martin,SalmanOL,SalmanOPEX,SalmanOMEX,MafiAOP}. In particular,
Karbasi, et al.~\cite{SalmanOL} observed TAL in a disordered polymer optical fiber quite similar to the structure proposed by 
De Raedt, et al. and used it for image transport~\cite{SalmanNature}: the mean localized beam radius (localization length) determines 
the width of the imaging point spread function (PSF) and hence the quality of the imaging system~\cite{SalmanNature}. 
In this Letter, we focus our attention on obtaining the MA-PDFs for the TALOF of the form presented in Refs.~\cite{DeRaedt,SalmanOL} for 
two reasons: first, it is a problem of practical importance especially for imaging
applications as discussed in Ref.~\cite{SalmanNature}; and second, it will be more straightforward to convey these general 
ideas in the context of a simple example. However, we emphasize that the results obtained in this works on MA-PDF and its scaling 
and saturation behavior are quite broad and can be applied to the study of various TAL disordered systems.

In order to calculate the MA-PDF, we need to solve for the guided modes of TALOF. Full vectorial Maxwell equations for electromagnetic 
wave propagation in a z-invariant optical fiber are used to calculate the fiber modes~\cite{multiphysics2016version}. 
For each guided mode, the mode area $A_{\rm eff}$ is defined as the standard deviation of the normalized intensity 
distribution $I(x,y)$ of the mode according to 
\begin{align}
\label{eq:sigma2}
A_{\rm eff} = \mathcal{C} \iint\,dx\,dy\,[(x-\bar{x})^2+(y-\bar{y})^2]\,I(x,y),
\end{align}
where $x,y$ are the transverse coordinates across the facet of the fiber and
$(\bar{x},\bar{y})$ are the mode center coordinates 
\begin{align}
\label{eq:xbarybar}
(\bar{x},\bar{y}) = \iint\,dx\,dy\ {\bf r}_T\,I(x,y),\ \ {\bf r}_T=(x,y).
\end{align}
The mode intensity profile is normalized such that $\iint_{-\infty}^{+\infty} I(x,y) dxdy = 1$ and 
the constant $\mathcal{C}=6$ is chosen such that $A_{\rm eff}=D^2$ for the uniform intensity distribution over 
a square waveguide of side-width $D$. 

Figure~\ref{fig:ModeProfile}a shows a typical disordered refractive index profile used in the simulations and Fig.~\ref{fig:ModeProfile}b shows the mode intensity profile for one of the computed 
localized modes. The cross-section of the disordered fiber is chosen to be square-shaped to resemble the geometry of TALOF in Refs.~\cite{DeRaedt,SalmanOL,SalmanNature}. 
The computed mode is transversely localized due to the strong disorder in the transverse dimensions. On the other hand, in the absence of disorder, a guided mode covers nearly the entire 
cross-section of the fiber: Figs.~\ref{fig:ModeProfile}c and d, show a periodic refractive index profile, and a typical mode intensity profile.

In the rest of the paper, we will explore MA-PDF curves for TALOFs. For each MA-PDF curve, we have calculated all the guided modes 
(typically tens of thousands of modes) for at least ten individual realizations of the specific disordered configuration under study in order to take 
into account the statistical nature of the disordered fiber. We should emphasize that for a device-level application of TALOF, a strong disorder is desirable 
to ensure self-averaging and therefore, reliable performance of a single disordered device realization~\cite{SalmanOL, SalmanNature}.

\begin{figure}[h]
\centerline{
\includegraphics[width=\columnwidth]{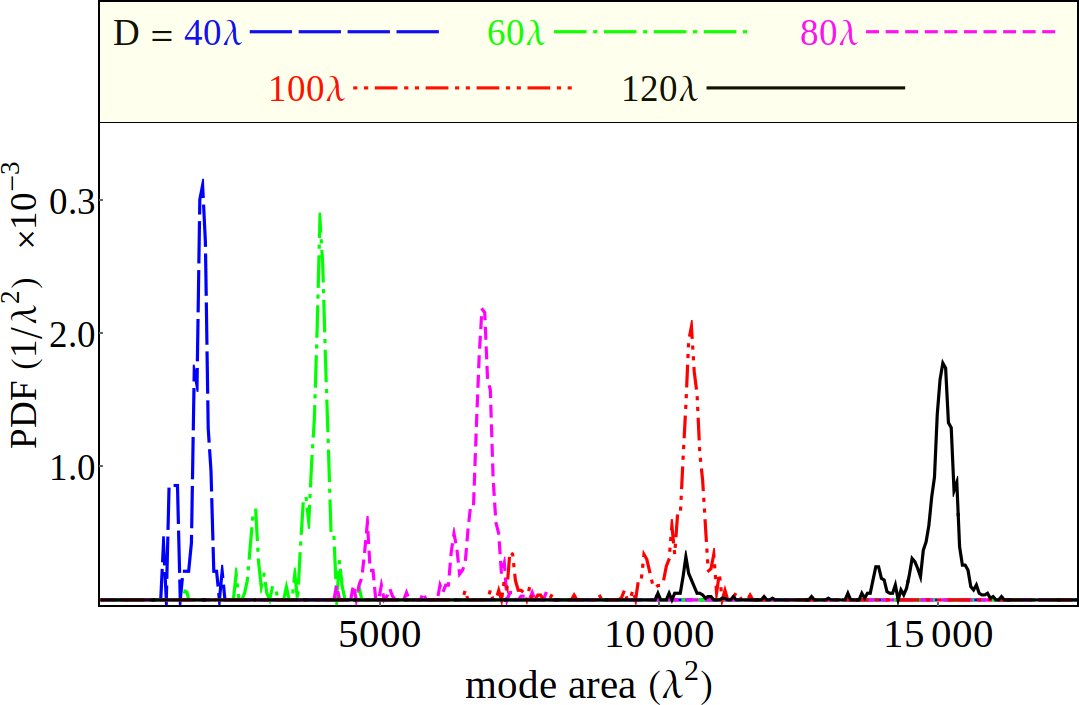}
}
\caption{\label{fig:PD1}
Mode-area PDF of a transversely periodic fiber formed by $N \times N$ number of unit cells for N = 20, 30, 40, 50, and 60 where the unit cell is a 
square of $2\lambda$ side-width. The average mode area of the transversely periodic fiber is solely determined by the transverse size of the fiber.}
\end{figure}
In the absence of disorder, guided modes of a conventional optical fiber are extended over the entire transverse dimensions of the fiber,
an example of which is shown in Fig.~\ref{fig:ModeProfile}(d);
total internal reflection caused by the index step formed at the interface between the fiber core and cladding is responsible for the overall transverse
wave confinement as it propagates freely in the third longitudinal direction. Therefore, the average transverse size of the guided modes scales proportionally to 
the cross-sectional area of the fiber. Figure~\ref{fig:PD1} shows the MA-PDF of transversely periodic fibers with $N \times N$ unit cells in the core.
In our notation, unless otherwise specified, each unit cell is a square with a side-width of $d=2\lambda$, so the area of each unit cell is $a=4\lambda^2$.
Therefore, the fiber core width of $D=$\,40, 60, 80, 100, and 120$\lambda$ in Fig.~\ref{fig:PD1} correspond to $N=$\,20, 30, 40, 50, and 60 unit cells in each direction. 
$n_1=1.58$, and $n_2=1.59$ are used in Fig.~\ref{fig:PD1} so the index step in the core of the fiber 
equals to ${\Delta n_{\rm core}}=n_2-n_1=0.01$. The horizontal axis is in units of $\lambda ^2$ and the vertical 
axis is in units of $1/\lambda ^2$ such that the total area under the PDF integrates to unity. The thickness of the cladding is $t = 5\lambda$ around the periodic 
core of the fiber and its refractive index $n_c$ is chosen to be equal to the lower core index $n_1$. Figure~\ref{fig:PD1} indicates that for the periodic fibers, the transverse size of the modes is solely determined by the size of the fiber core.
Therefore, the peak of the MA-PDF shifts towards larger values of mode area as the fiber becomes wider. 

For a disordered optical fiber, the scaling characteristics of the MA-PDF with respect to the transverse dimensions of the fiber are completely different from
the periodic case. Disorder-induced TAL localizes most of the guided modes across the transverse structure of the disordered
fiber and each mode provides a narrow guiding channel~\cite{ruocco2017disorder}.   
\begin{figure}[h]
\centerline{
\includegraphics[width=\columnwidth]{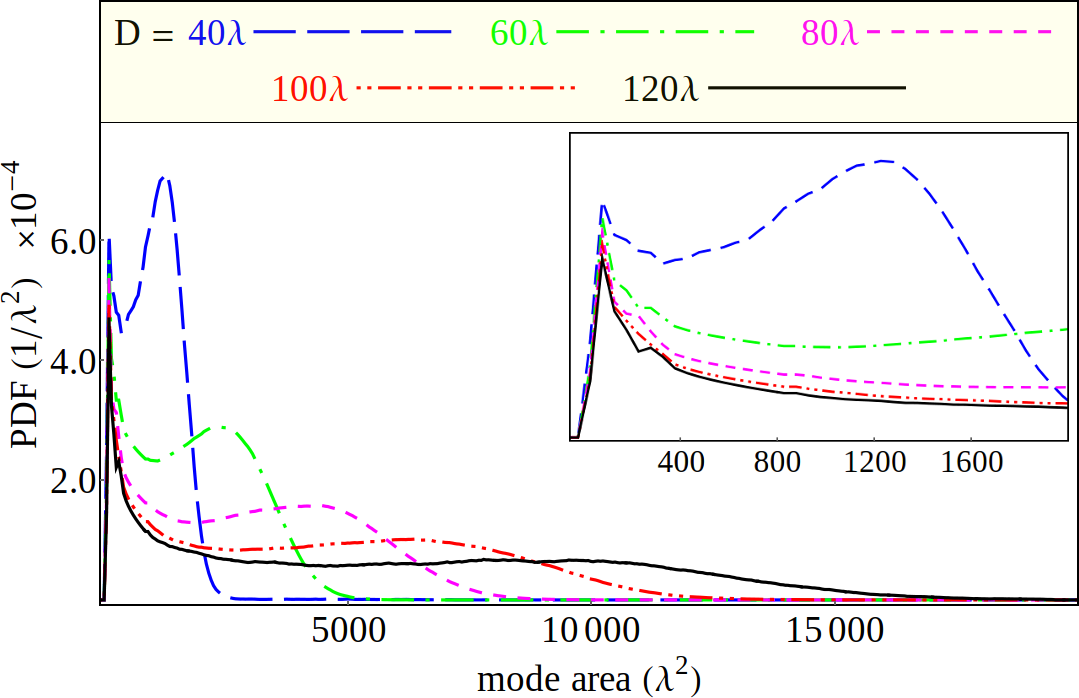}
}
\caption{\label{fig:PD2}
Mode-area PDF of a disordered fiber exactly the same as Fig.~\ref{fig:PD1},
except in the fact that refractive index of each unit cell is randomly chosen from the $(n_1, n_2)$ pair. 
The MA-PDF clearly represents presence of localized modes at $\sim 75\lambda^2$ and a long tail representative of 
guided extended modes. Moreover, saturation of the PDF in the highly localized region beyond $D_{\rm sat}\approx 100\lambda$ is clear in this figure. 
The inset is the magnified version of the localized peaks of the PDFs.}
\end{figure}
Figure~\ref{fig:PD2} shows the MA-PDF for a TALOF where all parameters involved in the disordered structure are exactly the same as Fig.~\ref{fig:PD1},
except in the fact that refractive index of each unit cell is randomly chosen from the $(n_1, n_2)$ pair with an equal probability. 
The MA-PDF shows a localized peak at a value less than $\approx 75\lambda^2$; therefore, the modes appear to be highly localized. The long tail of the PDF 
represents the guided extended modes that cover a large portion of the fiber cross section. 

In Fig.~\ref{fig:PD2}, the MA-PDF broadens as the number of cells is increased. The localized mode section below 1000$\lambda^2$ converges rapidly 
as a function of $D$ and remains nearly unchanged beyond $D_{\rm sat}\approx 100\lambda$. The tail of the distribution also converges albeit more slowly 
where the hump which is an artifact of the boundary gradually disappears and is replaced by a long decaying tail~\cite{PhysRevB.94.064201}.  

Convergence of MA-PDF beyond a critical number of cells indicates that $D_{\rm sat}$ can be considered as the effective transverse size of the 
disordered fiber beyond which the impact of the boundary on the localization properties is considerably reduced. Moreover, a disordered fiber with 
$D_{\rm sat}\approx 100\lambda$ can fully represent the behavior of a much wider fiber, hence significantly reducing the computational cost of simulating real-sized fibers. 
In order to see the saturation behavior of the MA-PDF more clearly, the inset shows a magnified version of the MA-PDFs, which is zoomed-in around smaller mode-area 
values.

\begin{figure}[h]
\centerline{
\includegraphics[width=\columnwidth]{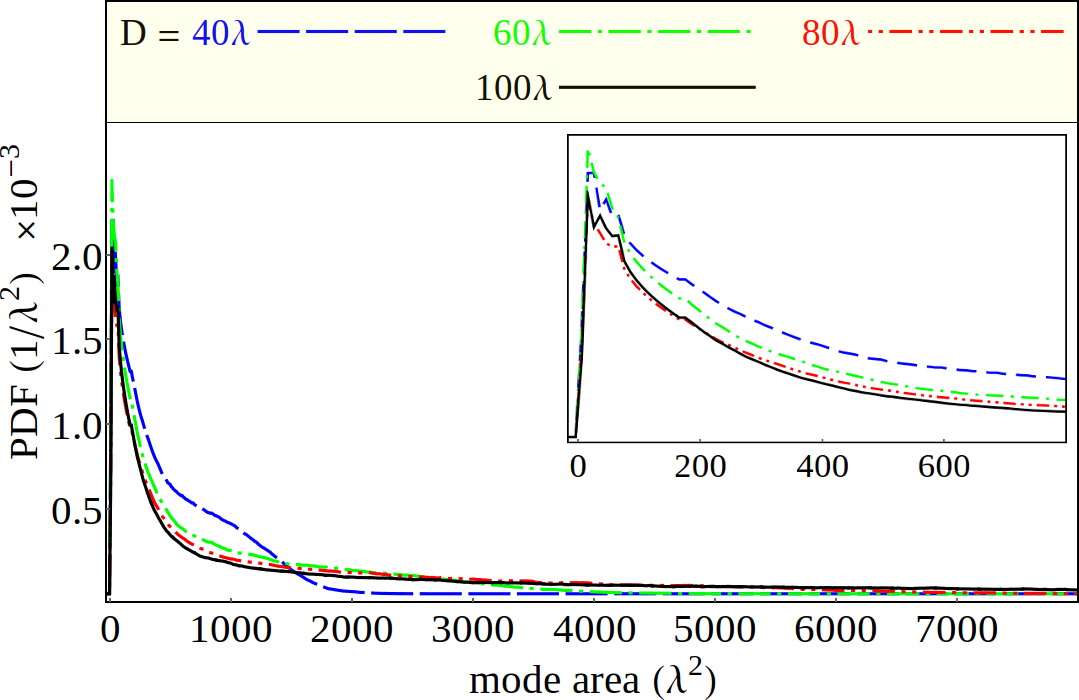}
}
\caption{\label{fig:PD3}
Mode-area PDF of a disordered fiber with exactly the same configuration as Fig.~\ref{fig:PD2} except for ${\Delta n_{\rm core}} = 0.1$.
Larger refractive index contrast in the disordered structure makes the PDF much narrower due to a much stronger localization. Inset is the magnified version of the localized peaks of the PDFs . Saturation of the PDF beyond $D_{\rm sat}\approx 60\lambda$ is clear in this figure.}
\end{figure}

The results shown in Fig.~\ref{fig:PD2} are related to a disordered fiber with an index difference of ${\Delta n_{\rm core}} = 0.01$.
If ${\Delta n_{\rm core}}$ is increased, the stronger transverse scattering must result in a stronger transverse localization and therefore smaller mode 
area values in the MA-PDF curve. A stronger transverse scattering is achieved by increasing the value of the index difference to  ${\Delta n_{\rm core}} = 0.1$ ($n_{1}=1.49$ and $n_2=1.59$). 
These values of $n_{1}$ and $n_2$ are chosen in accordance to the materials used in Ref.~\cite{SalmanOL} and the result is represented in Fig.~\ref{fig:PD3}. 
The peak values of MA-PDF curves occur at smaller mode area values ($\sim 15\lambda^2$), indicative of a stronger transverse localization. Moreover, the peaks are considerably narrower in Fig.~\ref{fig:PD3} compared with those in 
Fig.~\ref{fig:PD2}, because a stronger disorder also reduces the variance around the mean as discussed in Ref.~\cite{SalmanOPEX}, hence resulting in a more uniform
imaging PSF. Also, for the larger value of ${\Delta n_{\rm core}} = 0.1$, the convergence of the PDF occurs for considerably smaller disordered fibers ($D_{\rm sat} \approx 60\lambda$).

As an example for the utility of the concepts introduced here, we can tackle the important question of 
what the optimum value of the transverse scatterer size is to obtain the smallest imaging PSF. 
In the previous figures, it was assumed that the unit cells in the disordered core of the fiber are
squares with a side-width of $d=2\lambda$, in accordance to the TALOF reported in Ref.~\cite{SalmanOL}. 
Conceptually, one can imagine that if $d \ll \lambda$, the wavelength is much larger than the scatterer 
size and the medium appears as homogeneous to the wave. For $d \gg \lambda$, the system will be
in the geometrical optics limit. In either case, one expects a weak localization effect; therefore,
$d \sim \lambda$ is desired for maximum localization and the exact optimum value can be calculated from the relevant MA-PDF. 

\begin{figure}[htb]
\centerline{
\includegraphics[width=\columnwidth]{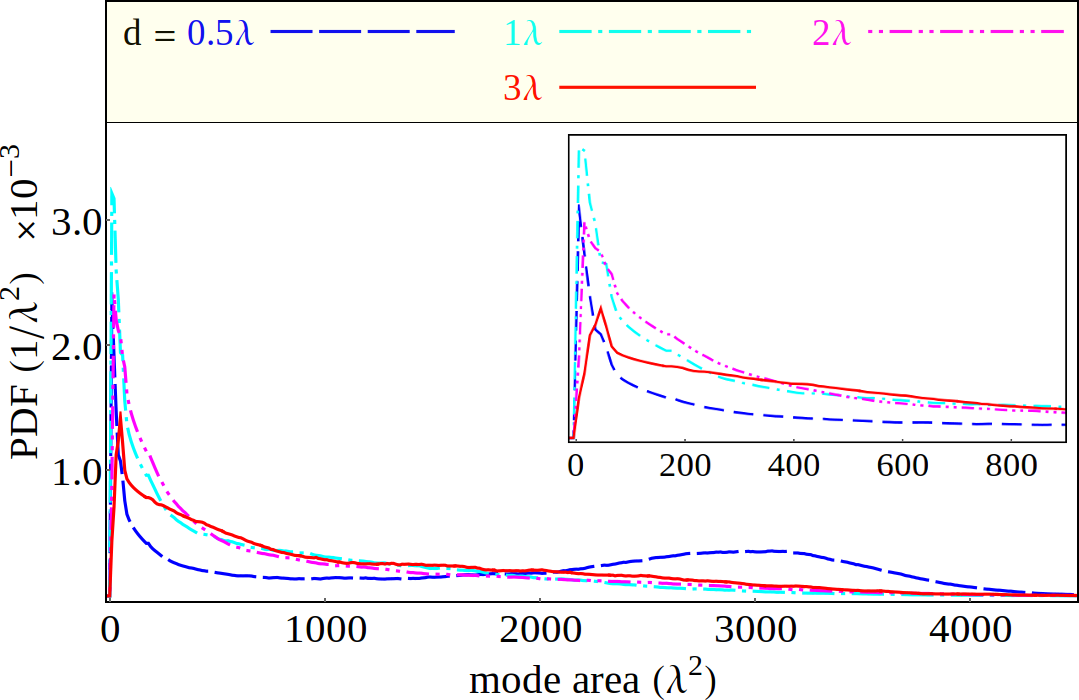}
}
\caption{\label{fig:PD4}
PDF of disordered fibers with $D=60\lambda$, ${\Delta n_{\rm core}=0.1}$,
and $d=\lambda/2,~\lambda,~2\lambda$, and $3\lambda$. $d \sim \lambda$ is desired 
for maximum localization strength.}
\end{figure}
In order to find the impact of the cell size, we compute in Fig.~\ref{fig:PD4} the MA-PDF curves in the saturation regime 
for TALOFs with several different values of cell size: $d=\lambda/2$, $\lambda$, $2\lambda$, and $3\lambda$.
In all cases, we assume that ${\Delta n_{\rm core}=0.1}$ and the fiber core width is $60\lambda$. 
The results seem consistent with the discussion above, where $d \sim \lambda$ is the desired vaue for maximum localization. Of course, the exact choice of 
the value of $d$ depends on a target function that must be calculated in a specific imaging scenario. It is not just
the localized modes that matter for imaging applications because the presence of the extended modes can also
reduce the contrast~\cite{KARBASI201372}; therefore, the optimization function must consider both the size of the PSF as well
as the contrast for the specific scenario. Figure~\ref{fig:PD4} clearly shows that for a given transverse size,
the MA-PDF in the extended mode region looks different for different values of $d$.

It is also important to point out that these results are not in contradiction with those reported in
Ref.~\cite{schirmacher2017right} where no dependence of the observed localization radii is found on the light 
wavelength. The results of Ref.~\cite{schirmacher2017right} are for averages of localization radii in the limit 
of infinite transverse fiber dimensions. How the MA-PDF can be averaged over the mode area in order to 
obtain the average localization radius of Ref.~\cite{schirmacher2017right} is beyond the scope of the 
present paper and will be dealt with in a future publication. 

In conclusion, a detailed statistical analysis based on MA-PDF is suggested as a powerful tool to 
study TAL of light, especially in disordered optical fibers. It is shown that the MA-PDF can be reliably
computed form structures considerably smaller than those used in experiments, hence substantially
reducing the computational cost to study TAL. The information contained in an MA-PDF can be used
to study and optimizes devices that operate based on TAL, e.g. in endoscopic image transport fibers~\cite{SalmanNature,ZHAO:17} and
directional fiber-based random lasers~\cite{BehnamRandomLaser}. The authors would like to point out that low-amplitude 
high-frequency oscillations of the MA-PDF cuvres presented in this paper have been smoothed out for clearer presentation
without compromising the results; however, original data can be obtained from the authors upon request. 

{\em The authors are thankful to the UNM Center for Advanced Research Computing (CARC) for providing access to
computational resources used in this work.}  
\bibliographystyle{apsrev4-1}
\providecommand{\noopsort}[1]{}\providecommand{\singleletter}[1]{#1}%
%
\end{document}